# Managing Complex Structured Data In a Fast Evolving Environment

Robert Smith
Secure Outcomes Inc.
2902 Evergreen Pkwy Ste 200
Evergreen, CO 80439 USA
rsmith@secureoutcomes.net

## ABSTRACT

Criminal data comes in a variety of formats, mandated by state, federal, and international standards. Specifying the data in a unified fashion is necessary for any system that intends to integrate with state, federal, and international law enforcement agencies. However, the contents, format, and structure of the data is highly inconsistent across jurisdictions, and each datum requires different ways of being printed, transmitted, and displayed. The goal was to design a system that is unified in its approach to specify data, and is amenable to future "unknown unknowns". We have developed a domain-specific language in Common Lisp which allows the specification of complex data with evolving formats and structure, and is inter-operable with the Common Lisp language. The resultant system has enabled the easy handling of complex evolving information in the general criminal data environment and has made it possible to manage and extend the system in a high-paced market. The language has allowed the principal product of Secure Outcomes Inc. to enjoy success with over 50 users throughout the United States[4].

## Categories and Subject Descriptors

D.2.11 [**Software Engineering**]: Software Architectures—*Domain-specific architectures*; H.2.3 [**Database Management**]: Languages—*Data description languages*; J.1 [**Computer Applications**]: Administrative Data Processing—*Government*

## General Terms

Design

## Keywords

Common Lisp, databases, domain-specific languages, specifications

## 1. INTRODUCTION

### 1.1 The LS1100

The LS1100 Executive Control System, or simply the *LS1100*, is a portable electronic fingerprinting, booking, transmission, and printing system intended for local, federal, and international law enforcement agencies. It has been developed by Secure Outcomes Inc. The LS1100 is fully certified by the Federal Bureau of Investigation (FBI) and is certified for digital transmission of fingerprint information to a number of U.S. states. The control software is written in over 100,000 lines of Common Lisp.

It is important to note that the LS1100 does *not* deal with biometrics, but rather the far more technically difficult *forensic livescan AFIS* ("automatic fingerprint identification system") domain. Biometric fingerprint systems are simpler commodity items designed to identify persons who want to be identified (for example, to get physical access through a door). AFIS systems such as the LS1100, in contrast, are designed to identify persons who are trying to hide their identity (for example, a murderer at a crime scene) with information that will stand up as *forensic evidence* in the court of law.

Fingerprint information that feeds the AFIS systems is far more difficult and complex to collect than data that feeds biometric systems and is covered by numerous state and federal standards, certifications, and regulations that constantly evolve. In general, biometric systems do not capture fingerprint information in a manner that is consistent with law enforcement fingerprint standards such as ANSI/NIST[3], FBI EFTS[5], and INTERPOL[1, 6], all of which the LS1100 formally follows.

The user work-flow of the LS1100 is simple. First, a detainee is brought to an agency's *booking station* where the system resides, and is prepared for fingerprint acquisition. Then the LS1100 is used according to the following general protocol:

1. Collect forensic quality, fully-rolled digital fingerprints from the system's built-in scan device.

2. Collect the detainee's personal information (e.g., height, weight, eye color, residence).

3. Print the criminal information on the local jurisdiction's fingerprint card.

4. Transmit the digital forensic information to the appropriate state or federal bureau.

Once this work-flow has been completed, an officer is able to later query the system for data, edit data, re-print cards, and re-transmit data.

Despite the relatively simple mode of operation from the user's point of view, the control software exhibits significant complexity due to the varying and evolving format and structure of the data. This paper deals with the design and implementation of the information aspects of the four steps above.

## 1.2 System Composition

The LS1100 control software is very complex; it is typically the only program running on top of the operating system. It compiles to an 85 megabyte executable and relies on a 500 megabyte collection of runtime files.

The source code is perhaps the most interesting. There are...

- ...over 105,000 lines of Common Lisp and 2000 lines of C in over 350 source files.

- ...over 50 packages.

- ...over 1300 exported symbols.

- ...over 2300 defined functions, 130 defined macros, 420 defined parameters, 130 defined variables, and 1000 defined constants.

- ...approximately 100 general purpose validators (described in section 4.4).

- ...over 500 *unique* defwidget-io forms (described in section 4.5).

- ...over 100 different classes and structures.

- ...over 60 defined CAPI[1] interfaces.

- ...about 50 third-party software packages.

The system is highly modular. There are approximately 40 discrete subsystems ranging in functionalities. For example, the system...

- ...contains over five methods of transmission of criminal data, including FTP, SFTP, Cisco VPN, and SMTP.

- ...can print on (with quarter-millimeter precision) over 30 types of fingerprint cards, some of which have up to 6 physical pages.

- ...contains over 5 context specific touchscreen keyboard types and about a dozen custom GUI widgets optimized for touchscreens.

---

[1] CAPI is LispWorks' portable GUI toolkit.

- ...includes a custom high-speed implementation of binary streams (independent of Lisp streams) with an implementation of the Advanced Encryption Standard (AES).

- ...includes many administrative tools, such as a database editor in CLIM, system stress tests, custom profilers, and performance measurement tools.

## 2. THE PROBLEM

The principal problem that arises from the four steps in section 1.1 is to create a coherent, unified model for the specification of data to be input (steps 1 and 2), and their automatic handling in printing and transmission (steps 3 and 4), suitable for a fast paced development environment. This comes from the fact that every state in the United States adheres to different data acquisition, formatting, printing, and transmission standards, but are all roughly derived from a single standard originally developed by the FBI. Moreover, all states and counties must be able to print on standard FBI fingerprint cards (the so-called "red" and "blue" cards), in addition to state- or county-specific fingerprint cards all of which have different formats.

## 2.1 Differing Constraints

Different legal jurisdictions impose different requirements on the data. For example, the state of Wisconsin requires a subject's name to comply with the following specification:

- The last name is required, and must consist of between 1 and 30 alphabetic characters.

- The first name is required, and must consist of between 1 and 20 alphabetic characters.

- The middle name is optional, but if present, must consist of between 1 and 20 alphabetic characters.

- The suffix is optional, but if present, must consist of between 1 and 4 alphabetic characters.

In contrast, the state of Arkansas requires 1–20 characters for the last name, 1–15 characters for the first name, and 1–3 characters for the suffix. Other data fields, such as hair color, are completely, incompatibly different.

## 2.2 Differing Printing/Transmission Formats

Each jurisdiction has its own methodology for printing on their jurisdiction-specific print media. For example, some states require a name be printed as `Doe, John S` while others require `John, S, Doe` — which has a different ordering of fields with a extra comma.

Transmission is similar, except transmissions are purely digital. There is a strict form as specified by the FBI in which data is transmitted, but varies slightly from jurisdiction to jurisdiction.

## 3. A UNIFYING DATA MODEL

In general, a piece of data is identified by and operated on by four pieces of specifying information:

1. The *name* $\nu \in N$: a representation relating a piece of data to some name. For example, $\nu$ might represent the concept of a subject's name or gender.

2. The *locality* $\ell \in L$: a representation of the jurisdiction that is collecting the data. For example, $\ell$ might represent the state of Arkansas.

3. The *usage* or *medium* $m \in M$: a representation of the data's *intent*. For example, $m$ might represent transmission.

4. The *index* of the data $k \in \mathbb{N}$: a natural number indicating which occurrence of the data is needed. For example, for a detainee with two known name aliases, the aliases would be identified by $k \in \{1, 2\}$.

The space of all quadruples $(\nu, \ell, m, k) \in N \times L \times M \times \mathbb{N}$ is called the *widget space* and elements of this space are called *widgets*. It only makes sense to talk about data when a widget is specified. It is convenient to think of the widget space as a space of triples $(\nu_k, \ell, m)$, where the first component is actually a vector of name-index pairs in $N \times \mathbb{N}$, which we will denote $\vec{N}$. We will use the latter convention, and denote the widget space by $\mathcal{W} = \vec{N} \times L \times M$.

### 3.1 Data Acquisition and Storage

The LS1100 system internally attempts to avoid handling of raw data by any means; almost all data is accessed through an opaque API which requires a widget. However, the raw data is required to be stored somehow, independent of the way the data is used. This suggests that the subspace of pairs $\vec{N} \times L$ define how the data is stored and retrieved. This is true in the LS1100 system; for each pair $(\nu_k, \ell)$, a *getter* $\gamma$ and *setter* $\sigma$ are defined. For a database $D \in \mathcal{D}$ and a datum of type $\Sigma$, we have

$$\gamma : \vec{N} \times L \times \mathcal{D} \to \Sigma \qquad (1)$$

$$\sigma : \vec{N} \times L \times \Sigma \times \mathcal{D} \to \Sigma \times \mathcal{D} \qquad (2)$$

In other words, the getter $\gamma$ takes an indexed name and locale, and queries the database for the datum. The setter $\sigma$ takes an indexed name, locale, and a datum, and inserts it into the database (forming a new database).

### 3.2 Formatters and Parsers

Each raw piece of data must be formatted, parsed according to the medium, and checked for errors in a process called *validation*. A *formatter* is a procedure which takes a datum and a widget $w$ and produces a string of characters representing that datum. A *parser* is the inverse; taking a string of characters and a widget and producing a datum.

The string of characters is always suitable for printing, screen display, and/or transmission.

### 3.3 Validation

Before any piece of data is parsed, it is validated by a validator. A *validator* is a function which takes a widget and a datum and returns a boolean representing its validity. A validator $v$ is usually a composition of smaller validators $v_1, \ldots, v_n$ which are generally composed via conjunction:

$$v_1 \circ v_2 : \mathcal{W} \times \Sigma \to \{\text{True}, \text{False}\} \qquad (3)$$
$$: (w, s) \mapsto v_1(w, s) \wedge v_2(w, s).$$

Other methods of composition, such as disjunction or negation, are possible and common.

It is important to note that validators are invariant to the widget's index.

## 4. PRACTICAL IMPLEMENTATION

All of the theory was realized in ANSI Common Lisp[2] as a domain-specific language.

### 4.1 Database Design and Implementation

The database was designed and implemented completely from scratch to accommodate the tight memory requirements of the LS1100, as well as to accommodate the flexibility required for the fast paced development style employed at Secure Outcomes Inc.

In the prototype of the LS1100 control software, SQLite was used together with an ORM between CLOS classes and SQL entities. After designing a database schema for the first state, it was quickly discovered that the traditionally static schema would not suffice because of the rapidly changing requirements. The database schema had to be redesigned in order to remove now invalid invariants as well as to encode new invariants, and the data in the databases had to be reorganized. Paired with limited versatility of data types[2] in SQL, it was considered that another database had to be created.

The new database design sacrificed the relational database model for a much more flexible key-value core augmented with a rich access and retrieval API. The database kernel was specifically designed to work with any serializable Common Lisp data type; almost any Common Lisp data type can be stored as values in the database, including hash tables, structures, classes, and functions. Name-index pairs, $\vec{N}$, with bounded indexes are implemented as vectors of values.

The core of the database is an on-disk, thread-safe hash table mapping Common Lisp keywords to data. Tables may be augmented with new keys and values at any time. Any queried key that doesn't exist will return a unique object called the *uninitialized value*. This serves as a type-safe null value indicating the lack of existence of either a key, or value associated with a key[3]. The uninitialized value can be manipulated and stored as well, and acts as an unboxed representation of Standard ML's `option` type or Haskell's `Maybe` type. Monadic behavior of the uninitialized value is achieved through a set of Common Lisp macros. See fig. 1 for a type-safe implementation of the uninitialized value.

Database queries are speed-optimized at the expense of memory; tables are cached in memory and written to disk at either a programmer-specified point or when the processor is available. The benefits far outweigh the costs, for the principal SQL table initialized to around 65 megabytes, while the database described initializes to around 1 kilobyte. A table filled with criminal data only increases the size by 1–5 kilobytes. (This excludes the forensic-quality fingerprint images, which are *not* stored in a table but stored separately on disk.)

---

[2] Often, data with no natural data type in SQL were "unparsed" to strings.

[3] It is actually used more generally to denote the lack of existence of data in other places, such as slots in structures.

## 4.2   Data

Common Lisp includes exemplary support for built-in data types, including flexible strings, a hierarchy of numerical types, and a selection of linear data structures. As discussed in section 4.1, almost all data types can be stored in the database[4], including programmer-defined structures and classes. By allowing a full breadth of types to be stored, data can typically be stored and retrieved in its most natural representation, without any need to transform it. The union of these types forms $\Sigma$ as described in section 3.

When adding a new data type to the LS1100 software, it typically suffices to create a structure representing the data, unless more control is needed. The new data structures should typically not place specific constraints on the size or format of the data, but rather just the types themselves. (Constraints are mandated by the validators. See section 4.4.) A type holding a name conforming to the specification of section 2.1 follows:

```
(defstruct name
  (last   "" :type simple-base-string)
  (first  "" :type simple-base-string)
  (middle "" :type simple-base-string)
  (suffix "" :type simple-base-string))
```

Since each name field is required by almost all state agencies and a default is provided, it is not necessary to anticipate an uninitialized value for any of the slot values.

Since Common Lisp provides mechanisms for compiling files to implementation specific binaries via COMPILE-FILE, and allows the loading of files via LOAD, it is possible to build a portable serialization protocol for most values. The initial version of the LS1100 data serializer was written in less than 25 lines of Common Lisp using the aforementioned standard functions, and has since been expanded for more intricate serialization and deserialization protocols.

## 4.3   Formatters and Parsers

Since data is usually stored in a fashion friendly to programmers and not users, we have implemented a protocol for converting between database data, or *internal data*, and data that was input or data to be read by a user, or *external data*.

Formatters, which map internal data to external data, are implemented as simple Common Lisp functions mapping data to strings[5]. For examples of formatters operating on simple date structures, see fig. 2.

There may be dozens of formatters for common data types, such as for names, to accommodate the needs of different agencies.

Parsers are a sort of inverse of formatters, mapping external data to internal data. Typically, a parser will take a string and convert it into a data structure suitable for storage. The string is usually guaranteed to be valid by the time a parser is called on it, thanks to the validation mechanism described in the next section. See fig. 3 for an example parser for date strings.

Since there exist many valid formatters and parsers associated with a single data type, we use the widget space to select which gets called for what kind of data. The specification of the widget space is described in section 4.5.

## 4.4   Validators

Validators are defined using a top-level macro DEFVALIDATOR. DEFVALIDATOR resembles DEFUN in many ways: it takes a name, argument list, and a body. However, validators aren't stored as the name's SYMBOL-FUNCTION, but rather stored in the symbol's property list as a validator. This allows validators to be stored with equivalent names as standard Common Lisp functions.

In the body of a validator, the symbol %INPUT is implicitly bound to the input that will be validated.

In practice, validators don't return a false value in the case of failed validation. Instead, a special VALIDATION-ERROR condition is signaled with information on the nature of the error, which is shown to the user. For example, consider the following two validators which check if a string is composed of numeric characters, and checks the length of a string, respectively.

```
(defvalidator numeric ()
  (loop
    :for c :across %input
    :unless (digit-char-p c)
    :do (validation-error %input
          "The character '~A' is not numeric" c)
    :finally (return t)))

(defvalidator length (min max)
  (let ((len (length %input)))
    (cond
      ((< len min)
       (validation-error %input
        "Length must be larger than ~D" min))
      ((> len max)
       (validation-error %input
        "Length must be smaller than ~D" max))
      (t t))))
```

---

[4]An exception includes native closures, which rely on run-time environment information.

[5]Though not necessarily. For fingerprint images, fingerprints are stored as compressed bitmaps and are converted to the binaries compressed with the wavelet scalar quantization algorithm, as mandated by the FBI.

**Figure 1: Simple implementation of the uninitialized value.**

```
(defun print-uninitialized (object stream depth)
  (declare (ignore depth))
  (print-unreadable-object (object stream)
    (princ "UNINITIALIZED")))

(defstruct (uninitialized (:print-function print-uninitialized))
  ;; no slots
  )

;;; DEFCONSTANT cannot be used because MAKE-UNINITIALIZED will not produce
;;; values that are EQL.
(defvar +uninitialized-value+ (make-uninitialized))

(defun uninitialized ()
  +uninitialized-value+)

(deftype maybe (&rest types)
  "A type representing a possibly empty value."
  `(or uninitialized ,@types))
```

**Figure 2: Sample formatters for date structures.**

```
(defun format-simple-date-short (date)
  "Format a date DATE in the American short way.

     EXAMPLE: 7/4/2010"
  (format nil "~A/~A/~A"
          (simple-date.month date)
          (simple-date.day date)
          (simple-date.year date)))

(defun format-simple-date-long (date)
  "Format a date DATE in the American long way.

     EXAMPLE: the fourth of July, 2010"
  (format nil "the ~:R of ~A, ~A"
          (simple-date.day date)
          (nth (1- (simple-date.month date))
               (list "January"   "February" "March"    "April"
                     "May"       "June"     "July"      "August"
                     "September" "October"  "November" "December"))
          (simple-date.year date)))
```

**Figure 3: Sample parser for dates.**

```
(defun parse-simple-date-fbi (str)
  "Parse the FBI-style date which may include optional slashes."
  (make-simple-date :year  (parse-integer (remove #\/ str) :start 0 :end 4)
                    :month (parse-integer (remove #\/ str) :start 4 :end 6)
                    :day   (parse-integer (remove #\/ str) :start 6 :end 8)))
```

Each of these signals an error condition to the user, specifying either which characters are the offending input, or what is wrong with the input in general. This is also why the `NUMERIC` validator does not use `EVERY` — we wish to know which character is invalid as well for a more constructive error message.

Several validators can be composed using a domain-specific language analogous to Boolean algebra. Logical conjunction of several validators is specified by the `VALIDATORS` form which coincides with (3). For most cases, conjunction is sufficient, since most validators are small and have a very simple semantic meaning.

In some cases, logical disjunction is useful for more complicated validations. As such, `OR` is defined inside of validator forms. For example, to create a validator which checks that the input must be between 3 and 7 characters, and must be either entirely alphabetic or entirely numeric, one constructs the following:

```
(validators
  (length 3 7)
  (or alphabetic
      numeric
      "Input must be alphabetic or numeric."))
```

The complete macro expansion of this can be found in fig. 4.

## 4.5   A Language for Widgets

A domain-specific language was developed in order to simplify the definition of objects in the widget space, and associating them with setters, getters, formatters, and parsers.

The core of the language is based around a macro `DEFWIDGET`, which takes the $(\nu, \ell)$ pair and a set of keys defining how a datum a widget refers to is handled.

The basic syntax follows:

```
WIDGET ::=
    (defwidget <WIDGET NAME> <LOCALE NAME>
      :index <INTEGER>            ; maximum index
      :getter <FUNCTION>
      :setter <FUNCTION>
      :input (<PARSER>*)
      :output (<FORMATTER>*)

FORMATTER ::=
    (<MEDIUM> <FUNCTION>)
  | (:default <FUNCTION>)

PARSER ::=
    (<MEDIUM> <FUNCTION> <VALIDATOR>)
```

The execution of this form will register the widget and the associated properties into the system.

## 4.6   Defaults

In practice, a lot of data properties are shared. For example, much of the criminal text data is stored as strings which can be displayed without modification. As such, the formatter is `IDENTITY` for all media of that widget. A particular piece of data called the "state ID number" follows this pattern.

```
(defwidget sid arkansas
  <...>
  :input ((:ls1100-entry 'identity
                          (validators alphanumeric
                                      (length 6 12))))
  :output ((:ls1100-entry    'string-upcase)
           (:lookup          'identity)
           (:fbi-criminal-249 'identity)
           (:fbi-applicant-258 'identity)
           (:transmission     'identity)))
```

We have found that it is more productive to simply provide a default, which is denoted by `:DEFAULT`.

```
(defwidget-io sid arkansas
  <...>
  :input ((:ls1100-entry 'identity
                          (validators alphanumeric
                                      (length 6 12))))
  :output ((:ls1100-entry 'string-upcase)
           (:default      'identity)))
```

Notice that the default can be overridden.

## 4.7   Inheritance

There is a lot of behavior sharing between (in the case of the United States) the FBI specifications, state specifications, and county specifications. This suggests a hierarchical notion of locales.

In almost all jurisdictions, the way a subject's date of birth is specified is the same. As such, we define a "common locale", symbolized by `COMMON`, which is the top of the locale tree.

```
(defwidget dob common
  <...>
  :input ((:ls1100-entry 'parse-date-fbi
                          (validators required
                                      date)))
  :output ((:ls1100-entry    'format-date-fbi)
           (:fbi-criminal-249 'format-date-card)
           (:fbi-applicant-258 'format-date-short)
           (:transmission     'format-date-fbi)
           (:DEFAULT          'format-date-card)))
```

In almost all jurisdictions, a date of birth will be formatted and parsed as specified. However, many states and local jurisdictions have specific unique fingerprint cards requiring certain kinds of formatting. So, for example, we define a widget for Arkansas-specific formatting.

```
(defwidget dob arkansas
  <...>
  :output ((:ar-arrest      'format-date-short)
           (:ar-supplemental 'format-date-short)))
```

Figure 4: Macro expansion of a `validator` clause.

```
(LAMBDA (#:TEXT-789583)
  (FUNCALL (VALIDATING-FUNCTION 'LENGTH 3 7) #:TEXT-789583)
  (FUNCALL (LAMBDA (#:INPUT-789600)
             (IF (EVERY #'NULL
                        (LIST (CATCH (LET ((#:OR-SUBFORM-RESULT789617
                                             CONDITIONS::*HANDLER-CLUSTERS*))
                                       (IF #:OR-SUBFORM-RESULT789617
                                           #:OR-SUBFORM-RESULT789617
                                           'CONDITIONS::NEW-IGNORE-ERRORS-HANDLER))
                                 (LET ((#:G789645
                                         'CONDITIONS::NEW-IGNORE-ERRORS-HANDLER))
                                   (DECLARE (DYNAMIC-EXTENT #:G789645))
                                   (LET ((CONDITIONS::*HANDLER-CLUSTERS*
                                           (CONS (LIST (CONS 'ERROR #:G789645))
                                                 CONDITIONS::*HANDLER-CLUSTERS*)))
                                     (FUNCALL (FUNCALL (SYMBOL-VALIDATOR 'ALPHABETIC))
                                              #:INPUT-789600))))
                              (CATCH (LET ((#:OR-SUBFORM-RESULT789672
                                             CONDITIONS::*HANDLER-CLUSTERS*))
                                       (IF #:OR-SUBFORM-RESULT789672
                                           #:OR-SUBFORM-RESULT789672
                                           'CONDITIONS::NEW-IGNORE-ERRORS-HANDLER))
                                 (LET ((#:G789700
                                         'CONDITIONS::NEW-IGNORE-ERRORS-HANDLER))
                                   (DECLARE (DYNAMIC-EXTENT #:G789700))
                                   (LET ((CONDITIONS::*HANDLER-CLUSTERS*
                                           (CONS (LIST (CONS 'ERROR #:G789700))
                                                 CONDITIONS::*HANDLER-CLUSTERS*)))
                                     (FUNCALL (FUNCALL (SYMBOL-VALIDATOR 'NUMERIC))
                                              #:INPUT-789600))))))
                 (ERROR 'VALIDATION-ERROR
                        :VALUE #:INPUT-789600
                        :MESSAGE "Input must be alphabetic or numeric.")
                 T)))
           #:TEXT-789583)
  T)
```

This Arkansas-specific definition inherits everything from the `COMMON` locale, and avoids duplicating the need to re-define higher jurisdiction cards (e.g., federal).

The `defwidget` macro does not actually specify inheritance since many defwidget forms refer to the same locale. Instead, a *locale inheritance tree* is defined via the

```
(add-locale!  child parent)
```

form, which constructs the inheritance tree by specifying a directed edge from `child` to `parent`.

Initialization of the locale inheritance tree starts by defining the root node `COMMON` which is the top of the lattice, and proceeds in a depth-first fashion:

```
;;; Root of locale tree
(add-locale! :common nil)

;;; United States
(add-locale! :united-states :common)

(add-locale! :colorado :united-states)
(add-locale! :park-county-co :colorado)
<...>

(add-locale! :minnesota :united-states)
(add-locale! :ramsey-county-mn :minnesota)
<...>

;;; Canada
(add-locale! :canada :common)

(add-locale! :nova-scotia :canada)
<...>
```

The locale inheritance tree is defined at runtime to allow hot-loading locale-specific patches and updates to the widget system (e.g., if a county requests a particular way of formatting a datum on their cards).

### 4.8 Application Programmer Interface

The principal API consists of two fusions of four functions:

$$\gamma \text{ fused with formatter} = \texttt{GET-AND-FORMAT-DATA}$$

and

$$\text{parser fused with } \sigma = \texttt{PARSE-AND-SET-DATA}$$

This fusion eliminates the need to explicitly retrieve data and consequently format it for a particular application, and removes the need to manually call a widget's validator, parsing function, and setter.

These functions are actually more than just a fusion of the getters and formatters (or parsers and setters); they do the extra work of actually finding and resolving the formatters and parsers. The resolution algorithm is as follows:

1. Check if the formatter/parser is defined at the current locale. If it is, then return it.

2. Check if a default formatter/parser is defined at the current locale. If it is, then return it.

3. Move to the current locale's parent.

   (a) If no parent exists, fail with "`No formatter/parser specified.`"

   (b) Otherwise, go to step 1.

### 4.9 Extensions In Production

The actual production code contains many extensions to the `DEFWIDGET` form.

Perhaps most importantly, it allows one to easily specify certain database tables in which the data is stored which has the benefit of automatically generating getters and setters for the programmer. These can always be overridden by specifying a getter and/or setter explicitly and also allows the widget system to be used for broader things, such as storing system settings.

Other extensions include the possibility to specify context-dependent headings for the data in the case the data should be presented to the user (e.g., for user queries or data summaries). It is also possible to specify on-line documentation, stored data type, and random data generators (useful for unit testing and generating random but valid test data).

## 5. ADVANTAGES AND ISSUES

The use of Common Lisp has allowed for the simple creation of a domain-specific language for specifying and validating data. Comparatively, in another language such as C, a special parser would be needed for the creation of a small language and extending the language would be difficult.

### 5.1 Common Lisp

The benefits of using Common Lisp extend beyond its exemplary metaprogramming facilities. Garbage collection — pioneered by Lisp systems but now common in many languages — makes managing resources convenient and suitable for long-running programs (the LS1100 is usually running for months at a time without restarting). Additional runtime stability comes from the fact that Lisp has been standardized since 1994 — and implementations have existed prior — allowing for implementations to mature and stabilize. Most behavior of the language is formally specified, and those areas of the language where behavior is ambiguous have been identified and informally resolved. Along with stable implementations come high-performance compilers. Since a large part of the LS1100 control software is dedicated to acquisition and analysis of high-resolution images, and since images are captured and displayed in real time on embedded hardware, it is necessary to have fast native code.

In general, the benefits of using Common Lisp for application development are well known. However, Secure Outcomes developers have also identified places where Common Lisp has fallen short, most importantly the lack of static types.

A static type system is one where the type of every variable and value is known at compile time, and the types of every interaction between variables and values are verified at compile time. In large applications such as ours, ensuring type correctness is almost an impossible task. The best that can be done is ensure runtime type checking is done. The problem with runtime type checking is that type errors are only found during the course of execution of the program, which can happen weeks or months later. When a type error is signaled, there's little chance of recourse when it is unanticipated and the user must restart the system entirely, which is often unacceptable when a detainee is required to be booked.

## 5.2 Custom Database

The advantages of using a custom database designed from the ground up for the application have largely been outlined previously. However, there are a few disadvantages.

One of the largest disadvantages is lack of maturity. Since the system is built from scratch, ensuring data integrity is a concern. In a law enforcement environment, loss of data can also be a loss of criminal evidence and records, and as such is completely unacceptable. Costly database backups must be made regularly in the event of a catastrophic database failure.

Another disadvantage caused by the flexibility of the possible data types stored is the difficulty in searching the data efficiently and the lack of query optimization — two problems largely solved in database research and implementations. Presently, the entire database must be walked for a general search. This problem is somewhat alleviated by providing specialized searches, such as name and alias searches, which use specially constructed indexes of the database.

## 6. CONCLUSION

Devising a domain-specific language in Common Lisp has been essential to the growth, rapid development, and delivery to the market of database interaction mechanisms for the LS1100 system. By modeling the data domain and how it is statically parameterized, and by developing simple methods for interacting with a database in this model, we were able to develop a flexible language that solves the problem of inconsistent and constantly changing specifications. The language takes advantage of natural inheritance and duplication in specifications, and was the catalyst for the ongoing success of the LS1100 livescan fingerprinting system.

Using Common Lisp to implement the LS1100 Executive Control System was a significant design decision. The decision proved to be correct — I believe that to implement the system in another language would have grossly extended the design, coding, and testing time required. In particular, the meta-programming capabilities uniquely provided by Common Lisp enabled the fast development of this complex commercial product that operates in its highly regulated environment.